\documentclass{article}
\usepackage{frascatiphys}
\usepackage{graphicx}
\usepackage{amsmath}
\usepackage{txfonts}
\usepackage[backend=biber,style=nature]{biblatex}
\begin{document}
\title{ 
CONSTRAINTS ON DARK ENERGY FROM THE ABUNDANCE OF MASSIVE GALAXIES
}
\author{
Paola Santini   \\
{\em INAF - Osservatorio Astronomico di Roma, via di Frascati 33, 00078 Monte Porzio Catone (RM), Italy} \\
Nicola Menci  \\
{\em INAF - Osservatorio Astronomico di Roma, via di Frascati 33, 00078 Monte Porzio Catone (RM), Italy} \\
Marco Castellano \\
{\em INAF - Osservatorio Astronomico di Roma, via di Frascati 33, 00078 Monte Porzio Catone (RM), Italy} \\
}
\maketitle
\baselineskip=11.6pt
\begin{abstract}
This conference proceedings paper provides a short summary of the constraints presented by Menci et al. (2020) and Menci et al. (2022) to dynamical dark energy models. 

Dynamical dark energy (DDE) models have been proposed to address several observational tensions arising within the standard $\Lambda$ cold dark matter ($\Lambda$CDM) scenario. Different DDE models, parameterized by different combinations of the local value of the equation-of-state parameter $w_0$ and its time derivative $w_a$, predict different maximal abundance of massive galaxies in the early Universe. We use the observed abundance of massive galaxies already in place at $z\gtrsim 4.5$ to constrain DDE models. To this aim, we consider four independent probes: (i) the observed stellar mass function at $z\sim 6$ from the CANDELS survey; (ii) the estimated volume density of massive haloes derived from the observation of massive, star-forming galaxies detected in the submillimeter range at $z\sim 5$; (iii) the rareness of the most massive system detected at $z\sim 7$ by the SPT survey; (iv) the abundance of massive ($M>10^{10.5}M_\odot$) galaxies at $z \sim 10$ as inferred from early JWST observations. Our probes  {\it exclude} a major fraction of the DDE parameter space that is allowed by other existing probes.  In particular, early JWST results, if confirmed, are in tension with the standard $\Lambda$CDM scenario at a 2$\sigma$ confidence level.
%presently allowed (or even favored) by other existing probes.
\end{abstract}
\baselineskip=14pt

\section{Introduction}
The current theory of structure formation envisages all cosmic structures to form from the collapse and the growth of initially tiny density perturbations of the dark matter (DM) density field in a universe characterized by an accelerated expansion. 
 Such an acceleration indicates that the dominant component (with density parameter $\Omega_\Lambda \simeq 0.7$) of the cosmic fluid must be composed of some form of dark energy (DE), with equation-of-state parameter $w=p / \rho \leq -1/3 $. 
 Although the nature of such a component remains unknown, the simplest model assumes DE to be connected with the vacuum energy, the so-called cosmological constant, with  $w=-1$. When coupled with the assumption that DM is composed of nonrelativistic particles at decoupling, such a scenario leads to the $\Lambda$ cold DM ($\Lambda$CDM) standard cosmological model  \cite{peebles93}. 

While measurements of the Cosmic Microwave Background (CMB) have provided a first, strong confirmation of such a scenario, tensions have recently emerged (\cite{menci20} and references therein) and  have stimulated an extended effort toward
the investigation of more complex cosmological models. One
of the simplest physical alternatives is  a DE with
a time-dependent equation of state, also called dynamical dark energy
(DDE)  (see \cite{huterer18} for a review).

The abundance of massive galaxies at high redshift
 constitutes a powerful probe for cosmological models. In fact, in the standard CDM
 scenario, large-mass DM haloes become progressively rarer with increasing redshift.
 The exponential high-mass
 tail of the mass function of DM haloes is
 expected to shift toward progressively smaller masses
 for increasing redshift (see, e.g., \cite{delpopolo07} for a review) at a rate that depends on
 the assumed cosmology. 
 Hence,  the comparison of  the predicted abundance of massive DM haloes at increasingly larger redshift with the observed abundance
 of galaxies with corresponding stellar mass $M_\ast$ provides increasingly strong constraints on the assumed
 cosmological framework.  
 Indeed, viable cosmological models must allow for an evolution of the initial density perturbations that is fast enough to match the abundance of massive galaxies observed to be in place at early epochs.

Under (extremely) conservative assumptions and considering different observables, we compare the maximal abundance of massive galaxies predicted in different DDE models at high redshifts with the measured abundance of the most massive systems observed to be already in place at the same redshifts. 
This conference proceedings paper summarizes the results that are presented and discussed in  \cite{menci20,menci22}.

\section{Method}

We compute the expected abundance of DM haloes, as a function of redshift $z$ and DM mass $M_h$, in different DDE models adopting the most conservative assumptions.  

We adopt the Sheth \& Tormen \cite{shethtormen01} mass function, within the Press and Schechter formalism. Besides being physically motivated and tested against N-body simulations for a variety of cosmologies, it has the most extended high-mass tail among the different proposed forms, hence representing the most conservative choice.  

The high-mass exponential cutoff  in the Sheth \& Tormen mass function is critically determined by the cosmic expansion rate and by the growth
factor, which depend on the equation of state of DE. For the latter, we use the CPL parameterization \cite{chevalier01,linder03} %to describe the cosmological evolution 
in terms of the scale factor $a$: 
\begin{equation}
w(a) = w_0 + w_a(1-a)
\label{eq:dde} 
\end{equation}
where the parameter $w_0$ represents the value of $w$ at the present epoch, while $w_a$ is its look-back time variation $w_a=-dw/da$. In the above parameterization, the standard $\Lambda$CDM cosmology corresponds to $w_0=-1$ and $w_a=0$. For each combination $(w_0 , w_a)$ , we can compute the expected number of DM haloes of mass $M_h$ as a function of redshift. We refer the reader to \cite{menci20} for a full description of the methodology and assumptions on the various involved cosmological parameters.
 
To compare these cosmological predictions on abundance of DM haloes with the measured  abundance of galaxies it is necessary to take into account the baryon physics. 
%
%We compare the predicted abundance of DM haloes with the abundance of galaxies measured from observations. 
However, baryonic effects are degenerate with cosmology in determining the expected  galaxy abundance. This can be bypassed by noticing that the ratio of galaxy baryonic components (stellar mass or gas mass) to DM halo mass has an absolute maximum at the cosmic baryon fraction $f_b$ ($f_b\simeq 0.16$, \cite{aghanim_planck18}). In fact, the observed abundance of galaxies with large mass in the baryonic component $M_b$ places a lower limit on the abundance of DM haloes with masses $M_h \geq M_b/f_b$. Such a constraint can be used to rule out cosmological models that do not allow for a sufficiently rapid growth of galactic DM haloes. In other words, since galaxies cannot outnumber their DM haloes, any  $(w_0 , w_a)$ combinations for which $\phi_{w_0,w_a}(M_h \geq M_b/f_b,z) \leq \phi_{obs}(M_b,z)$ can be excluded. 

Due to the exponential cutoff of the DM halo mass function and to its rapid redshift evolution, the highest masses at the highest redshifts put the most stringent limits. 
To adopt the most conservative assumptions, all our choices aim at maximizing $\phi_{w_0,w_a}(M_h,z)$ and minimizing $\phi_{obs}(M_b,z)$.

%  Since galaxies cannot outnumber their DM haloes, the latter measurement constitutes a lower limit to our predictions. In other words, we can rule out any cosmological model for which $\phi_{w_0,w_a}(M_h,z) < \phi_{obs}(M_b,z)$, where $M_b$ is the total galaxy baryonic mass. 
%The baryonic fraction $f_b$ ($f_b\simeq 0.16$, \cite{planck18}) imposes a lower limit on the abundance of DM haloes with masses $M_h \geq M_b/f_b $.

%Since the DM halo mass cannot be directly measured, the latter must be inferred from the galaxy baryonic component.

 %Since galaxies cannot outnumber their DM haloes, the observed abundance of galaxies with large mass in the baryonic component $M_b$ places a lower limit to the abundance of DM haloes. 

%The baryonic fraction $f_b$ imposes a natural limit on the stellar mass formed from a DM halo of mass $M$, as it cannot exceed $M_b=f_b*M$. We can therefore  rule out any cosmological model predicting a number density of DM haloes $\phi(M \geq M_b/f_b,z) \leq \phi_{obs}(M_b,z)$

\section{Results}

We describe here the various observables considered to constrain the  $w_0-w_a$ parameter space and the relevant regions excluded.

\subsection{Stellar mass function at $z=6$ from the CANDELS survey} \label{sec:mf}

We first compare with the observed stellar mass distribution
of massive, distant galaxies. Since stellar mass is a time-integrated quantity, it is not much sensitive to the details of the star
formation history \cite{santini15} and can be  easily related to the DM
mass of the host halo. An extended wavelength
coverage is essential for estimating stellar masses from spectral
energy distribution (SED) fitting, while a  combination
of survey volume and depth is required to
measure the
abundance of distante, massive and rare galaxies. The CANDELS project \cite{grogin11,koekemoer11},
taking advantage of the
optical/near-infrared/mid-infrared imaging provided by 
HST, Spitzer, and VLT on almost 1000 arcmin$^2$ down to faint fluxes,  provides an ideal data set to base such a
measurement. Here, we use the 
mass function derived by Grazian et al. \cite{grazian15}, who used a
spectral-fitting technique to derive stellar masses for a galaxy
sample with high-quality photometric redshifts based on the
CANDELS GOODS-South and UDS fields. 

We focus on their largest stellar mass bin (centered on
$M_\ast=8 \times 10^{10} M_\odot$, assuming a Salpeter initial mass function \cite{salpeter55}) at $z=5.5-6.5$. 
These high redshifts and  large masses ensure that the mass functions 
predicted by the different DDE models are in the full exponential regime, and are steep enough to make the comparison with the observed number density discriminant for the different DDE models. 

\begin{figure}[htb]
    \begin{center}
        {\includegraphics[scale=0.25]{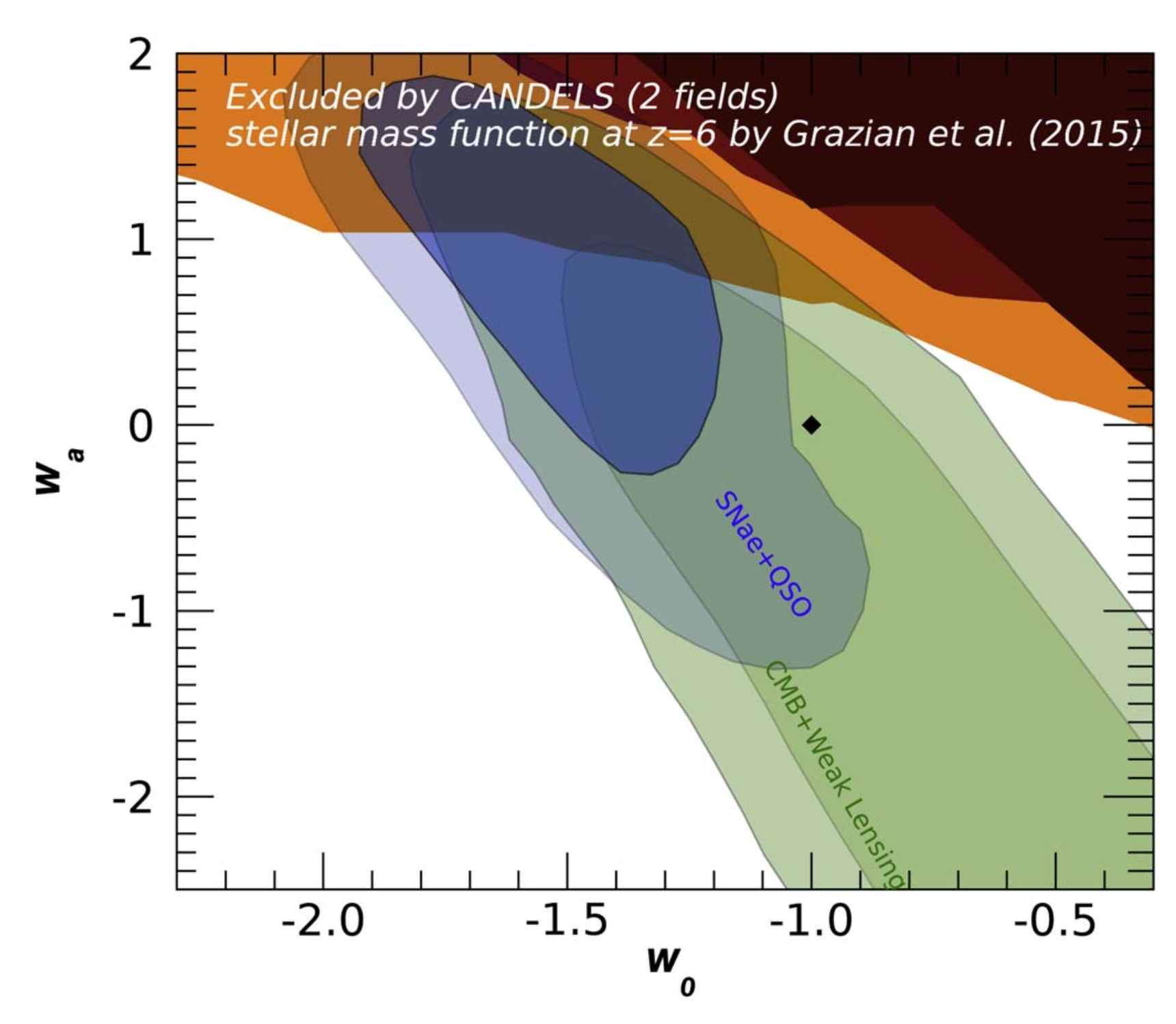}}\hspace{0.5cm}
        \caption{\it Exclusion regions at a 2$\sigma$ confidence level in the $w_0 - w_a$ plane derived from the observed CANDELS stellar mass function at $z \sim 6$ \cite{grazian15}. The brown, red, and orange regions correspond to the assumption of $F=1$, $F=0.5$ and $F=0.25$, respectively (see text). Our exclusion region is compared with the 2$\sigma$ and 3$\sigma$ contours allowed by CMB+weak lensing (green regions) and by the combination of the same data with the Hubble diagram of supernovae and quasars (blue region), derived from Figure 4 of \cite{risaliti19}. The black dot corresponds to the $\Lambda$CDM case ($w_0=-1$, $w_a=0$).}
\label{fig:mf}
    \end{center}
\end{figure}

We associate the stellar mass $M_\ast$ to the host halo DM mass $M_h$ using the relation $M_\ast=F f_b M_h$, where F describes the efficiency of baryon conversion into stars. We consider three cases: {\it (a)} the extremely unrealistic case  $F=1$, corresponding to a complete conversion;  {\it (b)} the more realistic case  $F=0.25$, as obtained from abundance matching techniques (e.g., \cite{behroozi18}) - this value however is derived assuming a $\Lambda$CDM halo mass function; and {\it (c)} the conservative value $F=0.5$, derived as a conservative upper limit on the star formation efficiency from hydrodynamical N-body simulations. 

Before comparing the predicted number density of DM haloes with observations, we rescale the observed volumes and luminosities  from a $\Lambda$CDM assumption to a generic cosmology through the factors $f_{Vol}=V_\Lambda/V_{w_0,w_a}$ and $f_{lum}=D^2_{L,w_0,w_a}/D^2_{L,\Lambda}$, respectively, where $V$ is the cosmological volume and $D^2_L$ is the square luminosity distance used to convert observed fluxed into luminosities, hence stellar masses. In summary, for each combination $(w_0 , w_a)$, we compare  the volume-corrected, observed abundance of 
galaxies $\tilde{\phi}=\phi_{obs} f_{Vol}$ with  stellar mass $M_\ast=8 f_{lum} 10^{10}M_\odot$ at $z\sim 6$ with the predicted number density of DM haloes with DM masses larger than $M_\ast/(F f_b)$, i.e. $\phi_{w_0,w_a}(M_h \geq M_\ast/(F f_b)$).   The confidence for the exclusion $P_{excl}$ of each considered DDE model is obtained from the probability distribution function $p(\tilde{\phi})$ as the probability that the measured abundance  $\tilde{\phi}$ is larger than number density predicted by  the model, i.e., $P_{excl}(w_0,w_a) = \int^\infty_{\phi_{w_0,w_a}}  {p(\tilde{\phi})d\tilde{\phi}}$.

We show in Figure~\ref{fig:mf} the region of the   $w_0-w_a$ parameter space excluded at a 2$\sigma$ confidence level (i.e., $P_{excl} \geq 0.95$) for $F=1$, $F=0.5$ and $F=0.25$. The exclusion region is overplotted on the regions allowed by CMB and weak lensing observations, and on the one derived by the combination of the same data with the Hubble diagram of supernovae and distant quasars \cite{risaliti19}. Our probe significantly restricts the region in the DDE parameter space allowed by other methods. In particular, we exclude an appreciable part of the region favored by the distant quasar method. Very similar results are obtained by comparing DDE predictions to the \cite{duncan14} mass function of CANDELS galaxies.

\subsection{Submillimeter detected massive galaxies at $z\sim 5$}

The  population of galaxies identified in rest-frame optical and ultraviolet is known to under represent the most
massive galaxies, which have rich dust content and/or old stellar
populations. These are, however, detectable at submillimeter (sub-mm)
wavelengths. Wang et al. \cite{wang19} performed detailed
sub-mm (870 $\mu$m) observations with ALMA of a sample of
IRAC–bright galaxies. They
detected 39 star-forming objects at $z>3$, which are unseen in
even the deepest near-infrared ($H$-band) imaging with the HST
(“$H$-dropouts”) and proved to be massive galaxies with
stellar mass extending up to $M_\ast \simeq 3 \times 10^{11}M_\odot$, with a median
mass $M_\ast \simeq 4 \times 10^{10}M_\odot$.

\begin{figure}[htb]
    \begin{center}
        {\includegraphics[scale=0.29]{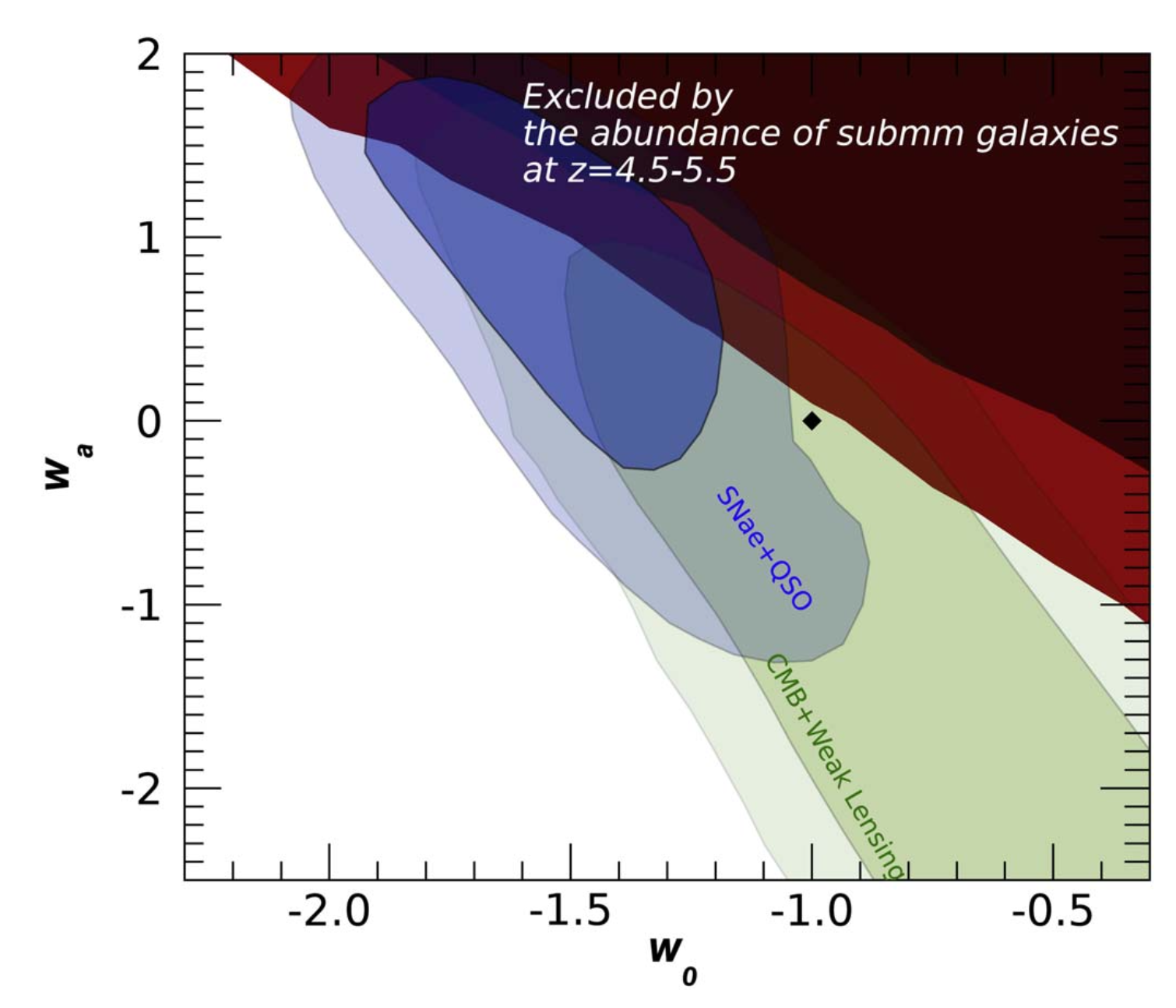}}\hspace{0.5cm}
        \caption{\it Exclusion regions at a 2$\sigma$ confidence level in the $w_0 - w_a$  plane derived from the observed abundance of luminous sub-mm galaxies at $z=4.5-5.5$ \cite{wang19}. The brown region corresponds to the assumption that the observed stellar masses are related to the DM mass through the baryon fraction $f_b$ ($M_\ast = f_b M_h$). The red region corresponds to adoption of the DM mass derived from the measured cross-correlation function of $H$-dropouts (see text).}
\label{fig:wang}
    \end{center}
\end{figure}

For such objects, we follow a procedure similar to that
explained in the previous section. We compute the number
density of galaxies with stellar masses in the bin $10.25 \leq \log(M_\ast/M_\odot) \leq 10.75$ (dominating the statistics of observed
objects) at redshifts $z=4.5–5.5$, and derive the corresponding
2$\sigma$ lower limit $\phi_{low}(M_\ast)=1.8 \times 10^{-5}$ Mpc$^{-3}$. 
To relate the
observed stellar mass $M_\ast$ to the DM mass $M_h$, we first adopt the
highly conservative assumption $M_h=M_\ast/f_b$ (i.e., $F=1$). 
The
comparison allows us to exclude (at a 2$\sigma$ confidence level) the
combinations $(w_0 , w_a)$ for which $\phi_{w_0,w_a} < \phi_{low}$. The result is 
shown as a brown exclusion region in Figure~\ref{fig:wang}.

Of course, the above approach is very conservative, since we assume that the whole baryonic mass is in stars, and that the baryon mass of DM haloes is related to the DM mass through the universal baryon fraction (no loss of baryons). However, the very fact that the objects are characterized by high star formation rates ($\gtrsim 200 M_\odot yr^{-1}$, \cite{wang19}) indicates that a sizable fraction of baryons is in the form of gas. Properly accounting for such a gas fraction would yield larger values of $M_h$ associated with the observed $M_\ast$ and, hence, tighter constraints. However, gas mass estimates for these objects are affected by extremely large uncertainties ($\sim$ a factor of 10) related to the uncertainties affecting the photometric redshifts and to all the assumptions required  to convert the sub-mm continuum flux into a gas mass. 
To bypass this and to derive more realistic constraints for DDE models, we analyze the clustering properties of the $H$-dropouts. We base our analysis on the procedure adopted by \cite{wang19}, who estimated the halo mass function from the measured correlation function, modified as described in \cite{menci20} to be adapted to a generic cosmology. We find that $M_h=10^{13}M_\odot$ constitutes a 2$\sigma$ lower limit for the value of the DM mass for any DDE model.  
The resulting exclusion region in the $w_0 - w_a$ plane is shown in red in Figure~\ref{fig:wang}. While the $\Lambda$CDM model remains marginally consistent with the observations, a much larger fraction of the  $w_0-w_a$ parameter space is excluded by the abundance of optically invisible, sub-mm galaxies at $z\sim5$.

\subsection{SPT0311-58 at $z=6.9$}

The most massive system detected at $z\geq 6$ is a far-infrared luminous object at redshift $z=6.9$  identified in the 2500 deg$^2$  South Pole Telescope (SPT) survey \cite{marrone18}. High-resolution imaging revealed this source (denoted SPT0311-58) to be a pair of extremely massive star-forming galaxies, with the larger galaxy (SPT0311-58W) forming stars at a rate of 2900 $M_\odot/yr$ and largely dominating over the companion. An elongated faint object seen at optical and near-infrared wavelengths is consistent with a nearly edge-on spiral galaxy at $z\simeq1.4$ acting as a gravitational lens for the source, with an estimated magnification $\mu=2$. 

The molecular gas mass was estimated from ALMA observations, both based on the CO luminosity and from a radiative transfer model \cite{strandet17}. We adopt the latter value of $M_{H_2}\simeq (3.1\pm 1.9) \times 10^{11}M_\odot$ as a baseline, since it is based on a detailed fit with a model built ad-hoc to study the  interstellar medium properties of this object and does not require assumptions on the conversion factor from the CO line to the $H_2$ mass \cite{marrone18}. 
To estimate the DM mass of the host halo of this galaxy, we assume that $M_h = (M_\ast + M_{gas})/f_b$, where $M_{gas}$ is the total gas mass. Since the stellar mass %$M_\ast$ 
cannot be directly measured due to the extremely faint optical emission of the galaxy (likely due to the large dust extinction), we can infer a lower limit on the stellar content from existing measurements of the molecular gas fraction $f_{H_2}=M_{H_2}/(M_\ast+M_{H_2})$. We consider the most conservative value $f_{H_2}=0.8$ measured on high-$z$ star-forming galaxies (see references in \cite{menci20}) and a more realistic value $f_{H_2}=0.4$ suggested by both theoretical models \cite{narayanan12} and observations \cite{tacconi13}. 
Assuming that $H_2$ constitutes 80\% of the gas mass  at high redshifts (an upper limit according to \cite{lagos11,lagos14}) leads to a DM mass ranging from $\simeq 2\times 10^{12}M_\odot$ to $\simeq 6 \times 10^{12}M_\odot$. An even larger DM mass would be consistent with the observations if the object lost the majority of its molecular gas content. 

\begin{figure}[htb]
    \begin{center}
        {\includegraphics[scale=0.35]{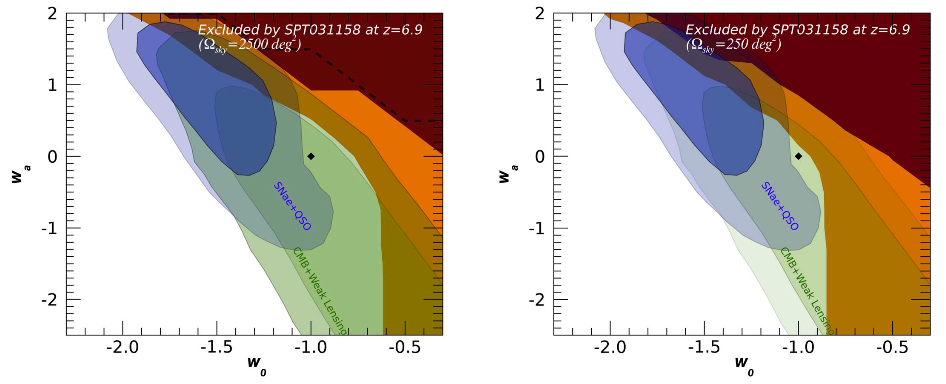}}\hspace{0.5cm}
        \caption{\it Exclusion regions at a 2$\sigma$ confidence level  in the $w_0 - w_a$ plane  for two different inferred DM masses of SPT0311–58: $2\times 10^{12}M_\odot$  (red area) and $2\times 10^{12}M_\odot$  (yellow area). The left panel assumes the full SPT survey area of 2500 deg$^2$ while the right one assumes 250 deg$^2$.
        }
\label{fig:SPT}
    \end{center}
\end{figure}

To estimate the rareness of such a system in all the considered DDE cosmologies, we compute the Poisson probability of finding such a massive object within the volume probed by the SPT survey, for different combinations $(w_0 , w_a)$. We follow the method in \cite{harrison13} adapted to a generic cosmology and take into account the uncertainties in the measured value of $M_{H_2}$ as described in \cite{menci20}. We consider the total area of the SPT survey, although the effective area is potentially much smaller. In fact, most of the objects in the survey are strongly lensed, indicating that a source must be gravitationally lensed to exceed the flux threshold for inclusion in the observations. For this reason, in the following we also consider the case of an effective area reduced by 1/10.
%
%For each combination $(w_0 , w_a)$, we compute the expected
%number of systems like SPT0311–58 detectable in the SPT
%survey. 
From the  rareness,  we compute the associated 2$\sigma$
exclusion regions in the $w_0-w_a$ plane.
 The result is shown in  Figure~\ref{fig:SPT}.

In the case $M_h = 6 \times 10^{11} M_\odot$, corresponding to the assumption of $f_{H_2} = 0.4$ for the $H_2$ gas fraction, a major portion of the $w_0-w_a$ plane is excluded, although the $\Lambda$CDM case ($w_0 = -1$, $w_a=0$) remains allowed. The excluded region
includes both the larger $w_a$ cases allowed by the quasar method (blue region) and the cases $w_0\geq -0.6$ allowed by the CMB + weak lensing results, which shows the potential impact of our results. Tighter constraints are obtained assuming an area of 250 deg$^2$, shown in the right panel of Figure~\ref{fig:SPT}.

\subsection{High-$z$ galaxies from early JWST results}

We finally exploit the very recent, early JWST results to derive even tighter constraints on DDE models. 
We compare the maximum stellar mass density $\rho_{max,w_0,w_a}(>M_\ast)$ allowed by different cosmologies with the  unexpected large stellar mass density measured by Labb\'e et al. \cite{labbe22}, who
%$\rho_{obs} (> M_\ast)$ measured by \cite{labbe22}. 
%They 
observed seven galaxies with $M_\ast \geq 10^{10}M_\odot$ at $7<z<11$. %, resulting in an unexpected large stellar mass density. . 
To derive the most stringent limits on cosmological models, we focus on their most massive bin, i.e. $M_\ast \geq 10^{10.5}M_\odot$ (derived assuming a conservative Chabrier initial mass function \cite{chabrier03}), in the redshift range $9<z<11$, yielding a stellar mass density of $\rho_{obs} \simeq 10^6 M_\odot/$Mpc$^3$. 
 
We compute the predicted maximal  stellar mass density for different $(w_0 , w_a)$ combinations. We  assume $M_h=M_\ast/f_b$ and adopt an even more conservative value for the baryon fraction of $f_b=0.18$ \cite{ade_planck16}. We rescale the observed stellar  mass density for the volume and luminosity correction factors to convert from $\Lambda$CDM to a generic DDE model as explained in  Sect.~\ref{sec:mf}. We derive the proper confidence level for exclusion for each considered cosmology by calculating the probability that $\rho_{max,w_0,w_a}(>\bar{M_\ast}) < \rho_{obs}(>\bar{M_\ast})$, where $\bar{M_\ast}=10^{10.5} M_\odot$. We run a Monte Carlo simulation to account for the observational uncertainties, assigning an errobar of 0.5 dex to the stellar mass to account for systematics related with the SED fitting procedure \cite{santini15}.

\begin{figure}[htb]
    \begin{center}
        {\includegraphics[scale=0.39]{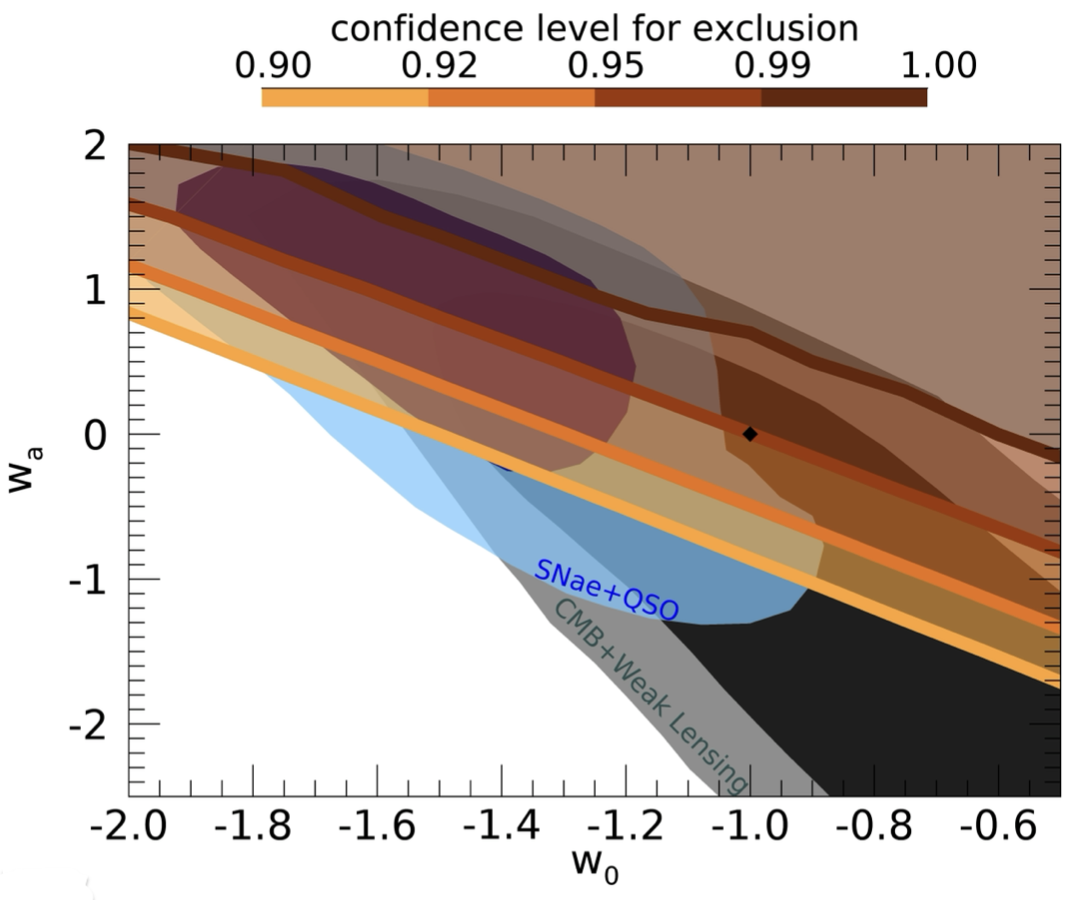}}\hspace{0.5cm}
        \caption{\it Exclusion regions in the $w_0-w_a$ plane derived from the
 observed stellar mass density at $z = 10$ \cite{labbe22}. The
 excluded regions above each coloured line correspond to different
 confidence levels shown in the upper bar. Our exclusion regions are compared with the 2$\sigma$ and 3$\sigma$ contours allowed by CMB+weak
 lensing (grey and dark-grey regions) and by the combination of the
 same data with the Hubble diagram of supernovae and quasars (blue
 regions), derived from Fig.~4 of \cite{risaliti19}. }
 \label{fig:jwst}
    \end{center}
\end{figure}

The resulting exclusion regions in the $w_0-w_a$ parameter space is shown in Figure~\ref{fig:jwst} for
different confidence levels, and compared with regions
allowed by existing probes. 
The $\Lambda$CDM case is excluded at almost 2$\sigma$ level,
while a major fraction of the parameter space is excluded with high confidence level. 
Our probe severely restricts the region in the DDE parameter space allowed by other methods, and exclude almost all the region favored by the distant quasar method.

\section{Discussion and Conclusions}

We have determined exclusion regions in the $w_0-w_a$ parameter space  of DDE models from the abundance of massive galaxies at early ($z>4.5$) epochs. 
Adopting the most conservative assumptions for the ratio between the observed baryonic component and the DM mass, as well as conservative choices for the cosmological parameters, we have derived robust constraints  that do not depend on the details of the baryon physics involved in galaxy formation. In addition, our results do not depend on the nature of the DM component \cite{menci22}.

All our probes exclude a major fraction of the parameter space favored by the quasar distances \cite{risaliti19}, including their best-fit combination $w_0\simeq-0.8$ and $w_a\simeq-1.5$. 
%They our results almost entirely rule out the quintessence models in which  the equation-of-state parameter $w$ decreases and initially $w> -1$ (“cooling” models), which occupy most of the region $w_0>-1$, $w_a>0$.
If confirmed, recent JWST observations are in tension with a $\Lambda$CDM scenario at a 2$\sigma$ confidence level. 
Our results leave open the possibility that the present tension in the value of $H_0$ between the values derived from Planck and those obtained from local luminosity distance measurements \cite{kamionkowski22} may be solved in DDE models, since the combinations $(w_0, w_a)$ that allow for the reconciliation of the different observations \cite{divalentino17,zhao17} include values outside our exclusion region. 
 
Our constraints will be greatly tightened when improved, reliable measurements of the actual baryon fraction in galaxies, and of the relative weight of each baryonic component, will be available. 
Increasing the statistics of high-redshift massive objects will also greatly tighten present constraints by reducing the uncertainties associated with the low abundance of these galaxies.

A critical issue is associated with the systematics dominating the error budget in the mass estimates, especially at high redshift. The advent of JWST has for the first time opened the possibility to measure the rest-frame optical emission at high redshift \cite{santini22}, which was previously possible, despite being subject to a high noise level, only for very few bright and  isolated sources detected with the Spitzer telescope. Early JWST observations have revealed their potential impressive impact in constraining cosmological models, as also shown by independent analyses \cite{boylankolchin22,lovell22}. %,gong22,maio22}. 
However, JWST observations are extremely new and may still be subject to revision. In particular,  the results of \cite{labbe22} were derived on the basis of the first set of calibrations released by STScI. A 10-20\% level revision in the NIRCam calibrations \cite{boyer22} is not expected to yield   to revisions of the stellar mass-to-light ratios of
 the targets large enough to affect significantly mass estimates and our
 conclusions. Nevertheless, we caution that the effect on the
overall shape of the galaxy SED (as well as the assumptions on the star
formation histories adopted in the SED-fitting procedure \cite{endsley22}) may reflect in a non-linear
way on the estimated physical parameters of some objects. Finally, we have just started studying in detail the physics of $z\sim 10$ galaxies, and cannot exclude that  the star formation process can be significantly different from the lower redshift Universe, where our models and estimate procedures are calibrated. 
In particular, as discussed by \cite{steinhardt22}, 
 the increase of gas temperatures in star-forming,
 high-redshift galaxies  could result in an increasing contribution of massive stars to the galactic light, which would yield significantly lower values for the stellar masses  compared to
 those measured by \cite{labbe22}.

%\section{References}
%References in bibliography must be referred to in the text by a 
%superscript number with a right-handed bracket. All references should 
%be organized to provide initials and last name of the first author, 
%volume (bold-face), page number, year (in brackets) of publication. 
%

\printbibliography
%\bibliography{biblio}
%\bibliographystyle{naturemag}
%\bibliographystyle{aa}
\end{document}